\documentclass[prd,superscriptaddress,amsfonts,amssymb,amsmath,showpacs,twocolumn]{revtex4-2}
\usepackage{bm}
\usepackage{amsfonts}
\usepackage{latexsym}
\usepackage{graphicx}
\usepackage{amsmath}
\usepackage{palatino}
\usepackage{mathpazo}
\usepackage{textcomp}
\usepackage{float}
\usepackage{booktabs}
\usepackage{dcolumn}
\usepackage{booktabs}
\usepackage{multirow}
\usepackage{hyperref}
\hypersetup{colorlinks,citecolor=blue}
\usepackage{amsmath}
\usepackage{xcolor}
\usepackage{orcidlink}
\usepackage{epsfig}
\usepackage{caption}
\usepackage{subcaption}
\usepackage{commath}
\usepackage{tabularx}
\def\jnl@style{\it}
\def\aaref@jnl#1{{\jnl@style#1}}

\def\aaref@jnl#1{{\jnl@style#1}}

\def\aj{\aaref@jnl{AJ}}                   
\def\apj{\aaref@jnl{ApJ}}                 
\def\apjl{\aaref@jnl{ApJ}}                
\def\apjs{\aaref@jnl{ApJS}}               
\def\apss{\aaref@jnl{Ap\&SS}}             
\def\aap{\aaref@jnl{A\&A}}                
\def\aapr{\aaref@jnl{A\&A~Rev.}}          
\def\aaps{\aaref@jnl{A\&AS}}              
\def\mnras{\aaref@jnl{Mon.~Not.~Roy.~Astron.~Soc.}}             
\def\prd{\aaref@jnl{Phys.~Rev.~D}}        
\def\prc{\aaref@jnl{Phys.~Rev.~C}}  
\def\prl{\aaref@jnl{Phys.~Rev.~Lett.}}    
\def\qjras{\aaref@jnl{QJRAS}}             
\def\skytel{\aaref@jnl{S\&T}}             
\def\ssr{\aaref@jnl{Space~Sci.~Rev.}}     
\def\zap{\aaref@jnl{ZAp}}                 
\def\nat{\aaref@jnl{Nature}}              
\def\aplett{\aaref@jnl{Astrophys.~Lett.}} 
\def\apspr{\aaref@jnl{Astrophys.~Space~Phys.~Res.}} 
\def\physrep{\aaref@jnl{Phys.~Rep.}}      
\def\physscr{\aaref@jnl{Phys.~Scr}}       
\def\commat{\aaref@jnl{Comm.~Math.~Phys.}}              
\def\science{\aaref@jnl{Science}}               
\def\cqg{\aaref@jnl{Classical Quant.~Grav.}}            
\def\jpcs{\aaref@jnl{JPCS}}                                     
\def\ijmpd{\aaref@jnl{Int.~J.~Mod.~Phys.~D}}                    
\def\grg{\aaref@jnl{Gen.~Relat.~Gravit.}}               
\def\rpp{\aaref@jnl{Rep.~Prog.~Phys.}}          
\def\npa{\aaref@jnl{Nucl.~Phys.~A}}        
\def\lrr{\aaref@jnl{Living Rev.~Rel.}}                   
\def\jcap{\aaref@jnl{J.~Cosmology Astropart.~Phys.}}    
\def\rmp{\aaref@jnl{Rev.~Mod.~Phys.}}   
\def\epjc{\aaref@jnl{Eur.~Phys.~J.~C}}

\allowdisplaybreaks[1]

\begin{document}

\color{black}       

\title{Observational constraints on freezing quintessence in a non-linear $f(R, L_m)$ gravity}

\author{Yerlan Myrzakulov\orcidlink{0000-0003-0160-0422}}\email[Email: ]{ymyrzakulov@gmail.com} 
\affiliation{Department of General \& Theoretical Physics, L.N. Gumilyov Eurasian National University, Astana, 010008, Kazakhstan.}

\author{M. Koussour\orcidlink{0000-0002-4188-0572}}
\email[Email: ]{pr.mouhssine@gmail.com}
\affiliation{Department of Physics, University of Hassan II Casablanca, Morocco.}

\author{A. Çalışkan}
\email[Email: ]{aylincaliskan@istanbul.edu.tr} 
\affiliation{Department of Physics, Istanbul University, Istanbul 34134, Turkey.}

\author{E. G\"udekli\orcidlink{0000-0002-1174-6110}}
\email[Email: ]{gudekli@istanbul.edu.tr} 
\affiliation{Department of Physics, Istanbul University, Istanbul 34134, Turkey.}

\author{S. Muminov\orcidlink{0000-0003-2471-4836}}
\email[Email: ]{sokhibjan.muminov@gmail.com}
\affiliation{Mamun University, Bolkhovuz Street 2, Khiva 220900, Uzbekistan.}

\author{J. Rayimbaev\orcidlink{0000-0001-9293-1838}}
\email[Email: ]{javlon@astrin.uz}
\affiliation{Institute of Fundamental and Applied Research, National Research University TIIAME, Kori Niyoziy 39, Tashkent 100000, Uzbekistan.}
\affiliation{University of Tashkent for Applied Sciences, Str. Gavhar 1, Tashkent 100149, Uzbekistan.}
\affiliation{Urgench State University, Kh. Alimjan Str. 14, Urgench 221100, Uzbekistan}
\affiliation{Shahrisabz State Pedagogical Institute, Shahrisabz Str. 10, Shahrisabz 181301, Uzbekistan.}

\begin{abstract}

In this paper, we investigate the freezing quintessence scenario in late-time cosmic expansion using a non-linear $f(R, L_m)$ gravity model, $f(R,L_m)=\frac{R}{2}+L_m^\alpha$, where $\alpha$ is a free parameter. We consider a solution for this model using an appropriate parametrization of the scale factor, and then the model is constrained by observational datasets, including CC, Pantheon+ (SN), and CC+SN+BAO. Our analysis yields results aligning closely with observational data. The Hubble parameter, deceleration parameter, matter-energy density, and EoS parameter of our model exhibit expected trends over cosmic time, supporting its physical validity. Furthermore, the model demonstrates consistency with the $\Lambda$CDM model in late times, displaying freezing behavior in the $\omega - \omega'$ plane and stability against density perturbations. Our findings suggest that the modified $f(R, L_m)$ gravity model is a credible approach to describing the universe's accelerating phase.


\end{abstract}

\maketitle

\section{Introduction}\label{sec2}

In recent times, various cosmological observations have bolstered the evidence for the late-time accelerated expansion of the universe. These observations include measurements from supernovae Type Ia (SN Ia) teams, large-scale structure (LSS) \cite{E11}, the Wilkinson microwave anisotropy probe (WMAP) \cite{D.N.}, the cosmic microwave background radiation (CMBR) \cite{C.L.,R.R.}, and baryonic acoustic oscillations (BAOs) \cite{D.J.,W.J.}. The convergence of these diverse datasets provides compelling evidence for the existence of dark energy (DE), a mysterious component driving the accelerated expansion of the universe. However, according to these cosmological observations, it is estimated that DE and dark matter (DM) together constitute approximately 95-96\% of the universe's composition, representing mysterious components that are not yet fully understood. In contrast, baryonic matter makes up only about 4-5\% of the total content of the universe \cite{Planck/2014,Planck/2016,Planck/2020}. Currently, general relativity (GR) is considered the most successful theory of gravitation, supported by numerous gravitational tests outlined in \cite{Corda/2009}. Despite its success, GR fails to offer a satisfactory solution to the enigma of DE and DM. This shortcoming implies that GR might not serve as the definitive gravitational theory to tackle all contemporary cosmological challenges. 

In the past few decades, various alternative approaches have been suggested in the literature to resolve the current cosmological dilemmas. Among these, the modified theory of gravity has emerged as the most promising contender for tackling the challenges presented by DE and DM in the universe. A leading approach to tackling the mystery of the dark content issue in the universe involves modifying GR through the $f(R)$ theory of gravity, where $R$ represents the Ricci scalar \cite{Buchdahl/1970,Nojiri/2011}. This theory proposes modifications to the Einstein-Hilbert action by replacing the Ricci scalar with a more general function $f(R)$, offering a framework to explain the accelerated expansion of the universe without the need for DE \cite{Staro/2007,Capo/2008,Chiba/2007,Oikonomou/2022}. Several other modified theories have been developed to address this issue \cite{Nojiri/2017}, such as the $f(R, \mathcal{T})$ theory, where $\mathcal{T}$ represents the trace of the momentum-energy tensor \cite{Harko/2011,Koussour_fRT1,Koussour_fRT2,Koussour_fRT3,Koussour_fRT4}, $f(T)$ theory, where $T$ represents the torsion scalar \cite{Paliathanasis/2016,Salako/2013,Myrzakulov/2011,Koussour_fT1}, $f(Q)$ theory, where $Q$ represents the non-metricity scalar \cite{Jim/2018,Nojiri/2024}, $f(R,G)$ theory, where $G$ represents the Gauss-Bonnet invariant \cite{Laurentis/2015,Gomez/2012}, and many others.

An extension of the $f(R)$ gravity theory incorporating an explicit coupling between the matter Lagrangian density $L_m$ and a generic function $f(R)$ was initially proposed in \cite{Nojiri/2004,Allemandi/2005,Bertolami/2007}. This matter-geometry coupling introduces an additional force orthogonal to the four-velocity vector, leading to non-geodesic motion of massive particles. The model was later generalized to include arbitrary couplings in both the matter and geometric sectors \cite{Harko/2008}. Recently, Harko and Lobo \cite{Harko/2010} introduced a more advanced generalization of matter-curvature coupling theories, known as $f(R, L_m)$ gravity. In this framework, $f(R, L_m)$ represents an arbitrary function of the Ricci scalar $R$ and the matter Lagrangian density $L_m$. This theory can be regarded as the most comprehensive extension of gravitational models formulated within the Riemannian geometric framework. This non-minimal coupling between geometry and matter leads to the non-conservation of the energy-momentum tensor, which has significant physical implications, including notable modifications to the thermodynamics of the universe, similar to those observed in $f(R,T)$ gravity \cite{Harko/2011}. This, in turn, produces an additional force perpendicular to the 4 velocities of particles in the geodesic equation of motion. Thus, the trajectories of test particles diverge from the geodesic paths that GR predicts \cite{THK-1}. The thermodynamic implications of $f(R,L_m)$ gravity have been extensively analyzed by Harko \cite{thermo}. The generalized conservation equations in $f(R,L_m)$ gravity are interpreted through the formalism of open thermodynamic systems, revealing a connection to irreversible matter creation processes. These processes, validated by fundamental particle physics, correspond to energy transfer from the gravitational field to matter constituents, consistent with the second law of thermodynamics. The study derives expressions for particle creation rates, creation pressure, entropy production rates, and temperature evolution laws for the newly created particles. Furthermore, the analysis shows the significant production of comoving entropy during cosmological evolution due to the geometry-matter coupling, underscoring the profound thermodynamic consequences of such interactions. The $f(R, L_m)$ gravity has attracted significant attention in the scientific community due to its capacity to effectively address a range of cosmological and astrophysical challenges \cite{THK-2,THK-3,V.F.-2,Harko/2014,Harko/2015}. Furthermore, $f(R, L_m)$ gravity does not obey the equivalence principle and is constrained by experiments conducted within the solar system \cite{FR,JP}. In recent times, there has been a notable surge in interest in investigating the intriguing cosmological implications of $f(R, L_m)$ gravity. An increasing number of studies are now focusing on various aspects of this model; for instance, refer to the cited Refs. \cite{GM,RV-1,RV-2,THK-7,THK-8}. Jaybhaye et al. \cite{Jaybhaye/2022} discussed the cosmological implications of $f(R, L_m)$ gravity. The authors investigated the behavior of the universe within the framework of the flat Friedmann-Lema\^{i}tre-Robertson-Walker (FLRW) metric, considering its impact on the evolution and structure of the cosmos. Myrzakulov et al. \cite{Myrzakulov/2023} explored the evolution of the effective equation of state (EoS) parameter in a non-linear $f(R, L_m)$ DE model. The authors employ Bayesian analysis of cosmic chronometers and Pantheon samples to constrain the evolution of this parameter. Myrzakulova et al. \cite{Myrzakulova/2024} investigated the DE phenomenon within the framework of $f(R, L_m)$ cosmological models. Specifically, they considered $f(R,L_m)=\frac{R}{2}+L_m^\alpha$ and $f(R,L_m)=\frac{R}{2}+(1+\alpha R)L_m$, where $\alpha$ is constant. The authors used observational constraints to analyze and understand the behavior of DE in these models. Koussour et al. \cite{Koussour/2024} explored the bouncing behavior in $f(R, L_m)$ gravity, focusing on phantom crossing and analyzing its implications for the energy conditions

In our study, we investigate the freezing quintessence scenario in late-time cosmic expansion using a non-linear $f(R, L_m)$ gravity model given by $f(R,L_m)=\frac{R}{2}+L_m^\alpha$, where $\alpha$ is a free parameter. We analyze the $\omega - \omega'$ plane, a tool proposed by Caldwell and Linder \cite{Caldwell_2005}, which is useful for distinguishing between different DE models based on their trajectories on this plane. This approach has been previously applied to quintessence DE models, leading to two distinct classes on the $\omega - \omega'$ plane. The region where $\omega' > 0$ and $\omega < 0$ corresponds to the thawing region, while the region where $\omega' < 0$ and $\omega < 0$ corresponds to the freezing region. It is noted that the expansion of the universe is more accelerating in the freezing region. In addition, we consider a solution for this model using a scale factor parametrization given by $a(t) = \sqrt[n]{\sinh(t)}$, where $n > 0$ is an arbitrary constant. The motivations behind this choice are presented in Sec. \ref{sec3}. Our methodology produces findings that closely match empirical data, which is constrained by datasets from observation, such as cosmic chronometers (CC), Pantheon SN, and BAO. The present manuscript is structured as follows. In Sec. \ref{sec2}, we introduce the action and fundamental formulation that govern the dynamics in $f(R, L_m)$ theory. We also derive the modified Friedmann equations corresponding to the flat FLRW universe. In Sec. \ref{sec3}, we adopt an $f(R,L_m)$ functional and then explore a solution for this model using a parametrization of the scale factor. In Sec. \ref{sec4}, we constrain the values of the model parameters using observational data from CC, SN, and the combined CC+SN+BAO dataset to ensure consistency with these observations. Further, in Sec. \ref{sec5}, we analyze the behavior of several parameters, including the Hubble parameter, deceleration parameter, matter-energy density, and the EoS parameter. Next, we analyze the $\omega - \omega'$ behavior in Sec. \ref{sec6}. Sec. \ref{sec7} is dedicated to examining the stability of our model. Finally, we present our key findings and conclusions in Sec. \ref{sec8}.

\section{$f(R, L_m)$ Theory and Cosmology}\label{sec2}

In this context, we examine the action for $f(R, L_m)$ gravity, which has been proposed in a previous study \cite{Harko/2010}. The action is expressed as
\begin{equation}\label{Action}
S= \int{\sqrt{-g}d^4x f(R,L_m)}, 
\end{equation}
where $f(R,L_m)$ is an arbitrary function of the Ricci scalar $R$ and the Lagrangian density of matter $L_m$, and $g$ is the determinant of metric tensor. In addition, we adopt the convention $8 \pi G = c = 1$, where $G$ and $c$ represent the Newtonian gravitational constant and the speed of light, respectively. By definition, the Ricci scalar curvature is expressed as $R= g^{\mu\nu} R_{\mu\nu}$, where $R_{\mu\nu}$ is the Ricci tensor, defined as $R_{\mu\nu}= \partial_\lambda \Gamma^\lambda_{\mu\nu} - \partial_\mu \Gamma^\lambda_{\lambda\nu} + \Gamma^\lambda_{\mu\nu} \Gamma^\sigma_{\sigma\lambda} - \Gamma^\lambda_{\nu\sigma} \Gamma^\sigma_{\mu\lambda}$. Here, $\Gamma^\alpha_{\beta\gamma}$ represents the Levi-Civita connection components, can be obtained as
\begin{equation}\label{4}
\Gamma^\alpha_{\beta\gamma}= \frac{1}{2} g^{\alpha\lambda} \left( \frac{\partial g_{\gamma\lambda}}{\partial x^\beta} + \frac{\partial g_{\lambda\beta}}{\partial x^\gamma} - \frac{\partial g_{\beta\gamma}}{\partial x^\lambda} \right).
\end{equation}

The field equation for $f(R,L_m)$ gravity [39] is derived by varying the action integral (\ref{Action}) with respect to the components of the metric tensor $g_{\mu\nu}$,
\begin{equation}\label{FE}
f_R R_{\mu\nu} + (g_{\mu\nu} \square - \nabla_\mu \nabla_\nu)f_R - \frac{1}{2} (f-f_{L_m}L_m)g_{\mu\nu} = \frac{1}{2} f_{L_m} \mathcal{T}_{\mu\nu},
\end{equation}
where $f_R \equiv \frac{\partial f}{\partial R}$, $f_{L_m} \equiv \frac{\partial f}{\partial L_m}$, $\Box \equiv \nabla ^{\mu}\nabla _{\nu}$; $\nabla _{\mu}$ is the covariant derivative, and $T_{\mu\nu}$ represents the energy-momentum tensor for matter, defined by 
\begin{equation}\label{EMT}
\mathcal{T}_{\mu\nu} = \frac{-2}{\sqrt{-g}} \frac{\delta(\sqrt{-g}L_m)}{\delta g^{\mu\nu}}.
\end{equation}

Now, applying covariant derivation to Eq. \eqref{FE}, we can express it as
\begin{equation}\label{CD}
\nabla^\mu \mathcal{T}_{\mu\nu} = 2\nabla^\mu ln(f_{L_m}) \frac{\partial L_m}{\partial g^{\mu\nu}},
\end{equation}
which shows the violation of energy-momentum conservation in $f(R, L_m)$ gravity. Thus, once the nature of the universe under study is known, one can construct viable cosmological models and verify the dynamics of the universe by appropriately choosing the spacetime metric.

In this paper, we investigate the FLRW metric, which describes a homogeneous and isotropic universe on large scales. This metric is based on the cosmological principle, which asserts that the universe is homogeneous and isotropic when viewed on a large enough scale and is a fundamental concept in cosmology. It is given by \cite{Ryden}
\begin{equation}\label{FLRW}
ds^2= -dt^2 + a^2(t)[dx^2+dy^2+dz^2],
\end{equation}
where $a(t)$ represents the scale factor that quantifies the cosmic expansion at a given time $t$. From the metric \eqref{FLRW}, the Ricci scalar is derived as
\begin{equation}
R= 6 ( \dot{H}+2H^2 ),
\end{equation}
where $H=\frac{\dot{a}}{a}$ represents the Hubble parameter, which characterizes the rate of expansion of the universe.

In addition, we consider the universe to be filled with a perfect fluid. In this context, a perfect fluid is a theoretical model used in cosmology to describe matter distribution. It is characterized by having no viscosity and zero thermal conductivity. The energy-momentum tensor of the perfect fluid is given by
\begin{equation}\label{EMT1}
\mathcal{T}_{\mu\nu}=(\rho+p)u_\mu u_\nu + pg_{\mu\nu},
\end{equation}
where $u^\mu=(1,0,0,0)$ are the components of the four-velocity of the perfect fluid, which satisfies the relationship $u^{\mu}u_{\mu}=-1$. 
In this context, $\rho$ represents the matter-energy density, while $p$ represents the isotropic pressure. Using the expression for the energy-momentum tensor, we can calculate its trace as $\mathcal{T}=g^{\mu \nu} \mathcal{T}_{\mu \nu}=3p-\rho$.

The modified Friedmann equations play a crucial role in describing the dynamics of the universe in $f(R, L_m)$ gravity. These equations govern the evolution of the scale factor $a(t)$. In the context of $f(R, L_m)$ gravity, the modified Friedmann equations take the form \cite{Jaybhaye/2022}
\begin{equation}\label{F1}
3H^2 f_R + \frac{1}{2} \left( f-f_R R-f_{L_m}L_m \right) + 3H \dot{f_R}= \frac{1}{2}f_{L_m} \rho,
\end{equation}
and
\begin{equation}\label{F2}
\dot{H}f_R + 3H^2 f_R - \ddot{f_R} -3H\dot{f_R} + \frac{1}{2} \left( f_{L_m}L_m - f \right) = \frac{1}{2} f_{L_m}p,
\end{equation} 
where the dots indicate derivatives with respect to cosmic time $t$. 

\section{Cosmological solutions} \label{sec3}

The system of field equations described in Eqs. (\ref{F1})-(\ref{F2}) consists of only two independent equations with four unknowns: $f$, $H$, $\rho$, and $p$. To fully solve the system and analyze the evolution of energy density and pressure, two additional constraint equations (extra conditions) are necessary. These constraints are essential for closing the system and obtaining a unique solution. Here, we use a specific functional form of $f(R, L_m)$ gravity, which is expressed as \cite{Jaybhaye/2022,Myrzakulov/2023,Myrzakulova/2024,Koussour/2024}
\begin{equation}\label{fRL} 
f(R,L_m)=\frac{R}{2}+L_m^\alpha,
\end{equation}
where $\alpha$ is constant. The model being considered is motivated by the functional form $f(R, L_m)= f_{1}(R) + f_{2}(R) G(L_m)$, which signifies a general coupling between matter and geometry \cite{Harko/2014}. For $\alpha=1$, the equations reduce to the standard Friedmann equations of GR. In the present scenario, with $L_m=\rho$ \cite{Harko/2015}, the modified Friedmann equations \eqref{F1} and \eqref{F2} yields,
\begin{equation}\label{F11}
3H^2=(2\alpha-1) \rho^{\alpha}, \\
\end{equation}
and 
\begin{equation}\label{F22}
2\dot{H}+3H^2=\left[(\alpha-1)\rho-\alpha p\right]\rho^{\alpha-1}.
\end{equation}

Now, one additional constraint remains. The scale factor of the universe is crucial in cosmology, particularly for comprehending the fate of the cosmos and the dynamics of late time. It is a key part of modern cosmological theories, explaining both the expansion of the universe and its relationship with DE. Barrow \cite{Barrow/1990} found the exact solution to the Einstein field equations by applying a simple parametrization of the pressure-density relationship. A scaling factor of the type $a(t) = \exp(At^f)$ is obtained by this parametrization, where $A > 0$ and $0 < f < 1$ are constants. Amirhashchi et al. \cite{Amirhashchi/2011} examined a scale factor given by $a(t) = \sqrt{t \exp(t)}$ to derive an exact solution of the field equations. Akarsu et al. \cite{Akarsu/2014} considered a hybrid expansion law for the scale factor, i.e. $a(t) = t^{\alpha} \exp(\beta t)$, which is a product of power-law and exponential functions, to fully solve the field equations. Odintsov and Oikonomou \cite{Odintsov/2014} investigated the matter bounce scenario within the framework of loop quantum cosmology using $f(R)$ gravity, with the scale factor given by $a(t)=\left(\frac{3}{4}\rho_ct^2+1\right)^\frac{1}{3}$. In this study, the focus is on a special form of the scale factor proposed by Chawla et al. \cite{Chawla/2012} and used in \cite{Koussour_hyp,Nagpal/2019}, which is given by
\begin{equation}
\label{hyp}
    a(t)=\sqrt[n]{\sinh (t)},
\end{equation}
where $n>0$ is an arbitrary constant. 

By using Eq. (\ref{hyp}), we obtain the Hubble parameter $H(t)$ as
\begin{equation}
\label{Ht}
    H(t)=\frac{\dot{a}}{a}=\frac{\coth (t)}{n}.
\end{equation}

To enable a meaningful comparison between theoretical results and cosmological observations, we express the time variable $t$ in terms of the redshift $z$ using $a(t)=\frac{1}{1+z}$ (where $a_0=1$) and Eq. (\ref{hyp}), yielding,
\begin{equation}
    t(z)=\sinh ^{-1}\left[\left(\frac{1}{z+1}\right)^n\right].
\end{equation}

Now, using the equation above, we can express the Hubble parameter in terms of redshift as
\begin{equation}
\label{Hz}
    H(z)=\frac{H_{0}}{\sqrt{2}}\sqrt{(1+z)^{2 n}+1}.
\end{equation}

By setting $z = 0$ in Eq. (\ref{Hz}), we conclude that $H(0) = H_0$, where $H_0$ denotes the present value of the Hubble parameter. So, we can replace the derivatives with respect to time with derivatives with respect to redshift using the relation $\frac{d}{dt}=-H(z)(1+z) \frac{d}{dz}$. Thus, the time derivative of the Hubble parameter can be expressed as
\begin{equation}
\label{dt}
\frac{dH}{dt}=-H(z)(1+z) \frac{dH(z)}{dz}.
\end{equation}

From Eqs. (\ref{Hz}) and (\ref{dt}), we get the expression for the time derivative of the Hubble parameter $H(z)$ in terms of the redshift $z$ as $\dot H=-\frac{H_{0}^2 n}{2} (1+z)^{2 n}$. Furthermore, the dynamics of the model given in Eq. (\ref{Hz}) are entirely determined by the model parameters ($H_0$, $n$). In the following part, we investigate the evolution of cosmological parameters by constraining these parameters ($H_0$, $n$) using available observational datasets.

\section{Data and methodology} \label{sec4}

In this section, we evaluate the validity of the parametrization of the scale factor by confirming its agreement with recent observational data. We incorporate a range of observational data, including the CC dataset, Pantheon+ sample of SN Ia dataset, and BAO dataset. For data analysis, we employed an MCMC (Monte Carlo Markov Chain) technique using the publicly available emcee package \cite{Mackey_2013}. This approach allowed us to constrain the model parameters ($H_0$, $n$), enabling an investigation of the posterior distribution of the parameter space. The analysis produced one-dimensional distributions illustrating the posterior distribution of each parameter and two-dimensional distributions showing the covariance between different parameters. These distributions were complemented by the $1-\sigma$ and $2-\sigma$ confidence levels. Here, we present the observational data used:

\begin{itemize}
\item \textbf{CC dataset}: The CC dataset offers a valuable method for directly constraining the Hubble rate $H(z)$ at various redshifts. In our analysis, we use 31 data points compiled from studies by \cite{Jimenez_2003,Simon_2005,Stern_2010,Moresco_2012,Zhang_2014,Moresco_2015,Moresco_2016}. The CC method entails the use of spectroscopic dating techniques on galaxies that evolve passively to estimate the age difference between two galaxies at different redshifts. This age difference allows for the inference of $\frac{dz}{dt}$ from observations, enabling the computation of $H(z) = -\frac{1}{1+z} \frac{dz}{dt}$. Therefore, CC data are considered highly reliable because they are independent of any specific cosmological model, do not require complex integration, and rely on the absolute age determination of galaxies \cite{Jimenez_2002}.

\item \textbf{SN dataset}: Recent observational findings regarding SN Ia have confirmed the presence of the accelerated expansion phase of the universe. Over the past two decades, there has been a notable increase in the amount of data collected from samples of SN Ia. In this study, we employ the Pantheon sample \cite{Scolnic_2018,Chang_2019}, one of the most extensive compilations of SN Ia data comprising 1048 points within the redshift range of $[0.01, 2.3]$.

\item \textbf{BAO dataset}: BAO studies the oscillations that originated in the early universe due to cosmological perturbations in the fluid consisting of photons, baryons, and dark matter. This fluid was tightly coupled through Thomson scattering. The BAO measurements include data from the Sloan Digital Sky Survey (SDSS), the Six Degree Field Galaxy Survey (6dFGS), and the Baryon Oscillation Spectroscopic Survey (BOSS) \cite{BAO1, BAO2, BAO3, BAO4, BAO5, BAO6}.

\end{itemize}

In our MCMC analysis, we employ the $\chi^2$ function for the combined CC+SN+BAO dataset as
\begin{equation}
\chi^{2}_{joint}=\chi^{2}_{CC}+\chi^{2}_{SN}++\chi^{2}_{BAO},
\end{equation}
where
\begin{equation}
\chi^{2}_{CC} = \sum_{i=1}^{31} \frac{\left[H(\theta_{s}, z_{i})-
H_{obs}(z_{i})\right]^2}{\sigma(z_{i})^2},
\end{equation}
\begin{equation}
\chi^{2}_{SN} =\sum_{i,j=1} ^{1048} \Delta \mu_{i} \left(
C_{Pantheon}^{-1}\right)_{ij} \Delta \mu_{j},
\end{equation}
and
\begin{equation} 
\chi _{BAO}^{2}=X^{T}C_{BAO}^{-1}X\,.
\end{equation}

For $\chi^{2}_{CC}$, the variable $i$ iterates over the 31 data points, each corresponding to a specific redshift $z_{i}$. $H(\theta_{s}, z_{i})$ represents the model-predicted Hubble parameter at redshift $z_{i}$, determined by the model parameters $\theta_{s}=(H_0,n)$. $H_{obs}(z_{i})$ denotes the observed Hubble parameter at redshift $z_{i}$, and $\sigma(z_{i})$ is the uncertainty associated with the observed value at that redshift. 

For $\chi^{2}_{SN}$, the variables $i$ and $j$ iterate over the 1048 SN Ia data points. $\Delta \mu_{i}=\mu_{\rm th}-\mu_{\rm obs}$ represents the difference between the distance modulus of the $i_{th}$ SN Ia data point and the corresponding theoretical prediction, while $C_{SN}^{-1}$ is the inverse covariance matrix of the Pantheon+ sample, which accounts for correlations between the SN Ia data points. Further, the calculated theoretical value of the distance modulus is given by $\mu _{th}=5log_{10}\frac{d_{L}(z)}{1Mpc}+25$, where $d_{L}(z)=c(1+z)\int_{0}^{z}\frac{dy}{H(y,\theta_{s} )}$ is the luminosity distance \cite{Planck/2020}. 

For $\chi^{2}_{BAO}$, $X$ is a vector that changes based on the specific survey under consideration, and $C_{BAO}^{-1}$ is the inverse covariance matrix for the BAO data \cite{BAO6}. The matrix $C_{BAO}^{-1}$ incorporates the uncertainties and correlations among the BAO data points. The transpose of $X$ is denoted by $X^{T}$. In addition, we use $\frac{d_A}{D_V}$, and these observables to constrain our model parameters by fitting them to the BAO data from various surveys: $d_A(z)=c \int ^z_0\frac{dz'}{H(z')}$ and  $D_V(z)=\left[\frac{d_A(z)^2 cz}{H(z)}\right]^{1/3}$. Here, $d_A$ is the angular diameter distance in the comoving coordinates, and $D_V$ is the dilation scale.

\begin{figure}[H]
\centering
\includegraphics[scale=0.59]{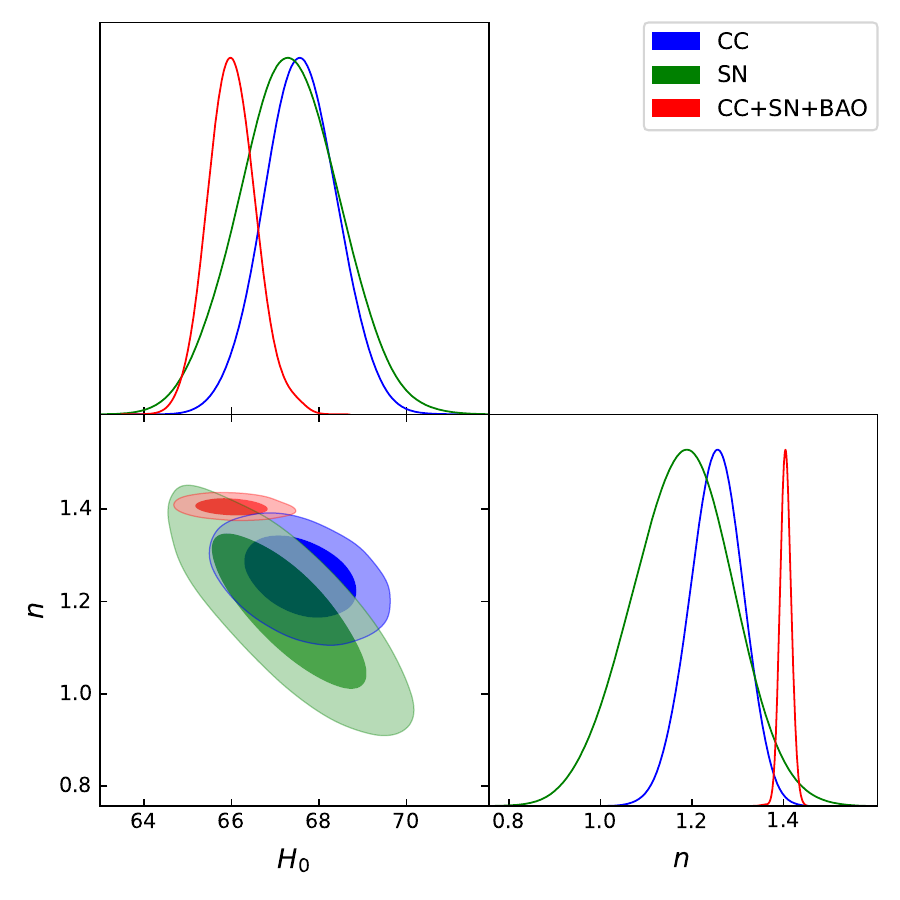}
\caption{The model parameters, namely $H_0$ and $n$, are represented by the confidence contours and posterior distributions.}
\label{F_Comb}
\end{figure}

Figs. \ref{H_err} and \ref{Mu_err} present a comparison between our model, and the widely-accepted $\Lambda$CDM model in cosmology, with $\Omega_{m0}=0.315 \pm 0.007$ and $H_0=67.4\pm0.5$ km s$^{-1}$ Mpc$^{-1}$ \cite{Planck/2020} considered for the plot. The figure incorporates the Hubble and Pantheon experimental results, comprising 31 and 1048 data points, respectively, along with their errors, facilitating a clear comparison between the two models. Further, Fig. \ref{F_Comb} shows the $1-\sigma$ and $2-\sigma$ likelihood contours for the model parameters $H_0$ and $n$ using the CC, SN, and CC+SN+BAO datasets. The results obtained through numerical computation using the MCMC method are summarized in Tab. \ref{tab}. In our analysis, we found that the value of $n$ for the combined dataset is concentrated around $n \sim 1.4$, primarily due to the significant influence of the BAO data. Specifically, with the BAO data alone, we determined $H_0 = 70.0^{+10}_{-9}$ and $n = 1.417^{+0.026}_{-0.025}$. This consistency across datasets justifies our focus on using the combined data for a more robust and comprehensive constraint on the model parameters. 

\begin{widetext}

\begin{figure}[h]
\centering
\includegraphics[width=18cm,height=6.5cm]{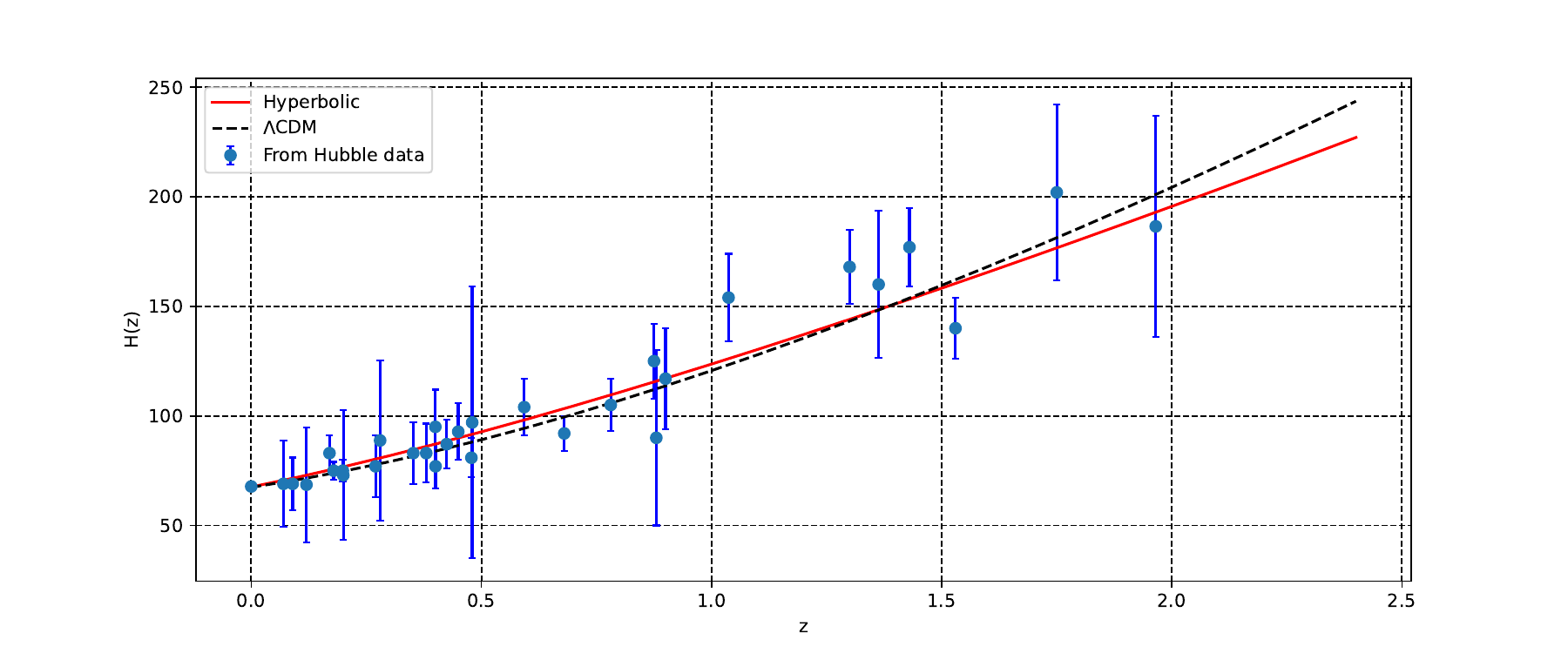}
\caption{The plot displays the Hubble parameter $H(z)$ versus redshift $z$ for our model (shown in red) and the $\Lambda$CDM model (shown in black dotted line), fitting nicely to the 31 points of the Hubble dataset, each accompanied by its respective error bars.}
\label{H_err}
\end{figure}

\begin{figure}[h]
\centering
\includegraphics[width=18cm,height=6.5cm]{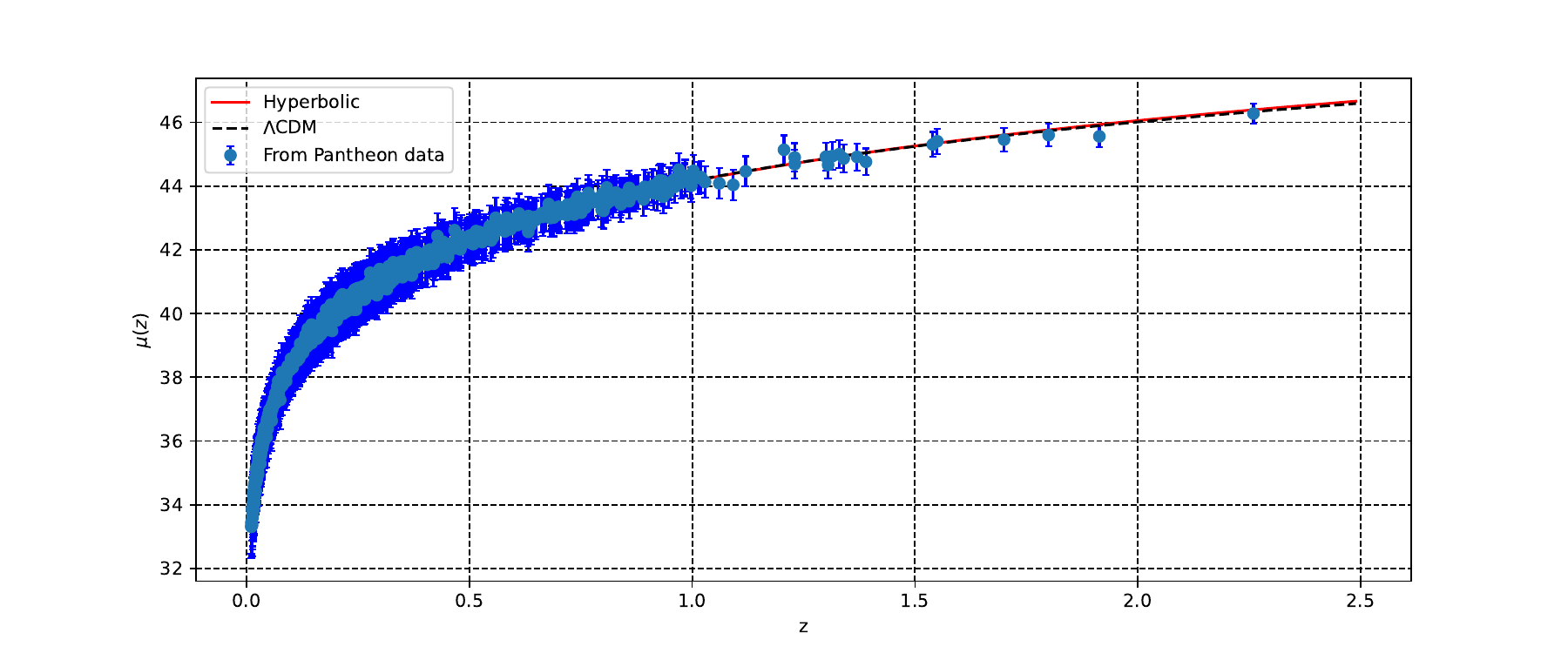}
\caption{The plot displays the distance modulus $\mu(z)$ versus redshift $z$ for our model (shown in red) and the $\Lambda$CDM model (shown in black dotted line), fitting nicely to the 1048 points of the Pantheon dataset, each accompanied by its respective error bars.}
\label{Mu_err}
\end{figure}

\end{widetext}

For statistical comparison of the hyperbolic parameterization of the scale factor with the $\Lambda$CDM model, we employ two model selection criteria: the Akaike information criterion (AIC) \cite{AIC1} and the Bayesian information criterion (BIC) \cite{AIC2}. The AIC is defined as \cite{AIC3, AIC4, AIC5}
\begin{equation}
    AIC\equiv \chi^2_{min}+2 k,
\end{equation}
and the BIC is defined as \cite{AIC4,AIC5,AIC6}
\begin{equation}
    BIC\equiv \chi^2_{min}+ k \log(N_{tot}),
\end{equation}
where $\chi^2_{min}=-2 \ln{(\mathcal{L}_{max})}$. Here, $k$ represents the number of parameters in the model, $\mathcal{L}_{max}$ is the maximum likelihood value for the analyzed datasets, and $N_{tot}$ denotes the total number of data points. We focus on the relative difference in information criterion (IC) values among the considered models. This difference, denoted as $\Delta IC_{model} = IC_{model} - IC_{min}$, compares each model's IC value to the minimum IC value among the competing models. According to Jeffreys' scale \cite{AIC7}, if $\Delta IC \leq 2$, the model is statistically consistent with the data's most favored model. A difference of 2 to 6 indicates moderate tension between the models, while a difference of 10 or more signifies significant tension. For these comparisons, we use the $\Lambda$CDM model as a reference and compare it with our model. We then constrain the parameters $H_0$ and $\Omega_{m0}$ of the $\Lambda$CDM model using the aforementioned datasets, with the outcomes displayed in Tab. \ref{tab}. These results are utilized in subsequent model selection analyses, the outcomes of which are also presented in Tab. \ref{tab}. From our analysis, the $\Delta IC$ values for the CC and SN datasets are both below 2. This suggests that our model agrees with both the $\Lambda$CDM model and the observational datasets. However, for the CC+SN+BAO datasets, the $\Delta IC$ value falls between 2 and 6, indicating a mild level of tension.

In the following section, we discuss the cosmological consequences of the observational constraints obtained. We examine the behavior of the Hubble parameter, deceleration parameter, matter-energy density, and the EoS parameter. These analyses are based on the best-fit values of the model parameters $H_0$ and $n$ constrained by the CC, SN, and CC+SN+BAO datasets. Since the model parameter $\alpha$ does not explicitly appear in the expression for the Hubble parameter, we fixed its value to study the evolution of matter-energy density and the EoS parameter. We used the value $\alpha = 1.33$, as constrained by observational datasets in Ref. \cite{Myrzakulov/2023}.

\begin{center}
\begin{table*}[!htbp]
		\begin{ruledtabular}
		    \centering
		    \begin{tabular}{c c c c c c c c c }
				Model & $H_0$ ($km/s/Mpc$) & $\Omega_{m0}$ & $n$ & $\chi^2_{min}$ & AIC & $\Delta$ AIC & BIC & $\Delta$ BIC\\
    	\hline
     & & & \textbf{CC} & & &\\
    $\Lambda$CDM & $67.81\pm 0.87$ &$0.327\pm 0.034$ &- & 14.51 & 18.51 & 0 & 21.38 & 0\\
    Hyperbolic & $67.6 \pm1.7$ &- &$1.25^{+0.11}_{-0.12}$ & 15.50 & 19.50 & 0.99 & 22.37& 0.99\\
    &&&&&&\\
    & & & \textbf{SN} & & &\\
   $\Lambda$CDM  &$72.33\pm 0.28$ & $0.383\pm 0.022$ & - & 1043.39 & 1047.39 & 0 & 1057.29 & 0\\
    Hyperbolic & $67.3^{+2.3}_{-2.2} $&- & $1.18^{+0.21}_{-0.22}$ & 1044.50 & 1048.5 & 1.11 & 1058.41 & 1.12\\
    &&&&&&\\
     & & & \textbf{CC+SN+BAO} & & &\\
    $\Lambda$CDM & $70.1 \pm 5.7$& $0.302_{-0.026}^{+0.021}$ & - & 1060.75 & 1064.75 & 0 & 1074.73 & 0\\
    Hyperbolic & $66.0_{-1.1}^{+1.2}$ & - & $1.405_{-0.024}^{+0.025}$ & 1066.05 & 1070.05 &5.3 & 1080.03 & 5.3\\

\end{tabular}
		   \end{ruledtabular}
     \caption{Summary of best-fit model parameters, statistical analyses, and information criteria for the CC, SN, and BAO datasets.}
		    \label{tab}
		\end{table*}
\end{center}

\section{Cosmological parameters}\label{sec5}

Studying cosmological parameters is crucial for understanding various aspects of the universe. These parameters, which describe the overall dynamics of the universe including its expansion rate and curvature, are of significant interest in explaining the universe's formation from its constituent elements such as baryons, photons, neutrinos, DM, and DE. In any viable physical model, these parameters play a critical role. In our work, we focus on fundamental parameters like the Hubble parameter, deceleration parameter, and EoS parameter within the framework of the parametrization of the scale factor in $f(R,L_m)$ gravity.

\subsection{Hubble parameter}

The Hubble parameter $H$ represents the rate at which the universe is expanding at a given time. Recent observational findings indicate that the Hubble parameter is decreasing as the universe evolves. In this study, we investigate the relationship between the Hubble parameter and redshift $z$ based on the constrained values of the model parameters, as shown in Fig. \ref{H_err}. The data clearly indicates that the value of the Hubble parameter decreases with the evolution of the universe, which is in agreement with observational findings. Specifically, we observe that $H$ increases as $z$ increases. Based on recent observational data from Planck collaborators \cite{Planck/2020}, the Hubble constant has been determined to be $H_0 = 67.4 \pm 0.5$ km s$^{-1}$ Mpc$^{-1}$. 
From Fig. \ref{H_err}, we obtain the value of the Hubble constant as $H_0=67.6^{+1.7}_{-1.7}$ km s$^{-1}$ Mpc$^{-1}$ for the CC dataset, $H_0=67.3^{+2.3}_{-2.2}$ km s$^{-1}$ Mpc$^{-1}$ for the SN dataset, and $H_0=66.0^{+1.2}_{-1.1}$ km s$^{-1}$ Mpc$^{-1}$ for the combined dataset. This indicates that the model is in complete agreement with the observational value of the Hubble constant.

\subsection{Deceleration parameter}

The deceleration parameter is a fundamental parameter that characterizes the expansion history of the universe. From Eqs. (\ref{Hz}) and (\ref{dt}), we get
\begin{equation} \label{dtH}
    \frac{\dot H}{H^2}=n \left(\frac{1}{(1+z)^{2 n}+1}-1\right).
\end{equation}

Therefore, the deceleration parameter for the model is obtained as
\begin{eqnarray} \label{qz}
    q(z)&=&-1-\frac{\dot{H}}{H^2}\nonumber\\&=&n-1-\frac{n}{(1+z)^{2 n}+1}
    \nonumber\\
&=&
    \begin{cases}
      n-1, & z\to \infty \\
      -1, & z\to -1.
    \end{cases}
\end{eqnarray}

\begin{figure}[h]
\centering
\includegraphics[scale=0.7]{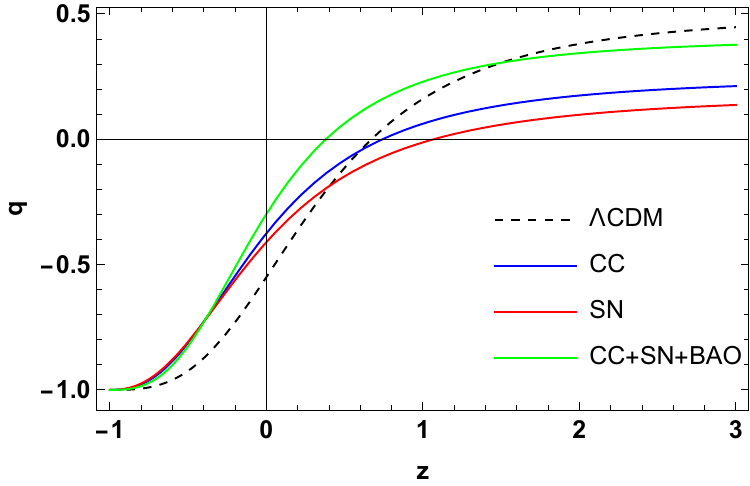}
\caption{Plot of deceleration parameter $q$ versus redshift $z$.}
\label{F_q}
\end{figure}

The sign of the Hubble parameter $H$ in the model indicates whether the universe is expanding or contracting, while its acceleration or deceleration is indicated by the sign of the deceleration parameter $q$. A positive $q$ signifies a decelerating model, while a negative $q$ indicates acceleration. It is noteworthy that current observations, such as SNe Ia and CMBR, tend to favor accelerating models with $q < 0$, although they do not conclusively support this scenario. From Eq. (\ref{qz}), it is evident that the deceleration parameter $q$ decreases monotonically from $n-1$ to $-1$. This implies that the universe's expansion transitions from deceleration in the early epoch (for $n>1$) to acceleration in the far future. Now, we present the plot of $q$ against redshift $z$ for this model in Fig. \ref{F_q}. Analyzing the figure, we observe the behavior of the deceleration parameter corresponding to the constrained values of the model parameters. We note that the deceleration parameter decreases as the redshift decreases, reaching more negative values. This negative value of the deceleration parameter indicates the accelerated expansion of the universe, which aligns perfectly with the observational data. Further, the present values of $q$ are determined to be $q_0 = -0.38^{+0.06}_{-0.06}$, $q_0 = -0.41^{+0.10}_{-0.11}$, and $q_0 = -0.30^{+0.01}_{-0.01}$ for the CC, SN, and combined datasets, respectively \cite{Cunha1,Cunha2,Nair,Koussour3,Koussour4,Koussour5}.

\subsection{EoS parameter}

Next, we examine the behavior of the EoS parameter $\omega$ in this model. The EoS parameter is one of the most important parameter which characterizes the properties of the energy density of the universe. It is defined as the ratio of the pressure of a substance to its energy density ($\omega=\frac{p}{\rho}$). From Eqs. (\ref{F11}) and (\ref{F22}), we get the EoS parameter as
\begin{equation}
    \omega=-1+\left(\frac{2-4 \alpha }{3 \alpha }\right)\frac{\dot{H}}{H^2}.
    \label{EoS}
\end{equation}

From Eq. (\ref{dtH}), we have
\begin{equation}
\label{w}
    \omega(z)=-1+\frac{2 (2 \alpha -1) n (1+z)^{2 n}}{3 \alpha  \left((1+z)^{2 n}+1\right)}.
\end{equation}

For different components of the universe, such as matter, radiation, and DE, the EoS parameter takes on different values, influencing the universe's evolution. For non-relativistic matter, such as baryonic and dark matter, the EoS parameter is approximately $\omega \approx 0$, indicating that the pressure is negligible compared to the energy density. Relativistic particles, such as photons and neutrinos, have an EoS parameter of $\omega = 1/3$, reflecting the fact that their pressure is one-third of their energy density due to their high speeds. DE, the mysterious component driving the accelerated expansion of the universe, is characterized by a constant or time-varying EoS parameter. A cosmological constant, often identified with DE, has $\omega = -1$ and exerts a negative pressure that counteracts gravity, leading to the observed acceleration. Quintessence is a form of DE with a time-varying EoS parameter, typically rolling down a potential energy field. It has $-1 < \omega < -1/3$ and behaves differently from a cosmological constant, leading to interesting cosmological dynamics \cite{quint}. Phantom energy is another type of DE with $\omega < -1$, which leads to even more rapid expansion, potentially resulting in a "Big Rip" scenario where the universe is torn apart by the increasing DE density \cite{phant}. Observational data, including measurements from SN Ia, CMBR, and BAOs, constrain the value of the EoS parameter for DE. Current constraints suggest that $\omega$ is very close to $-1$, indicating that DE behaves very much like a cosmological constant. However, small deviations from $\omega = -1$ are still being investigated to understand the nature of DE more fully.

\begin{figure}[h]
\centering
\includegraphics[scale=0.7]{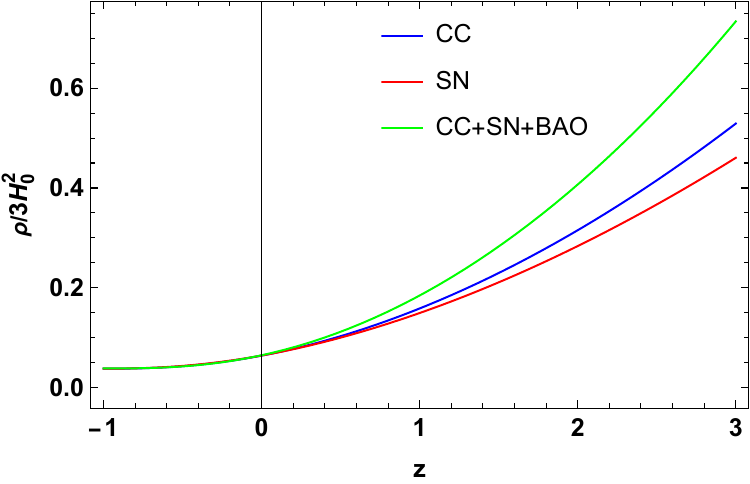}
\caption{Plot of matter-energy density $\rho$ versus redshift $z$.}
\label{F_rho}
\end{figure}

\begin{figure}[h]
\centering
\includegraphics[scale=0.7]{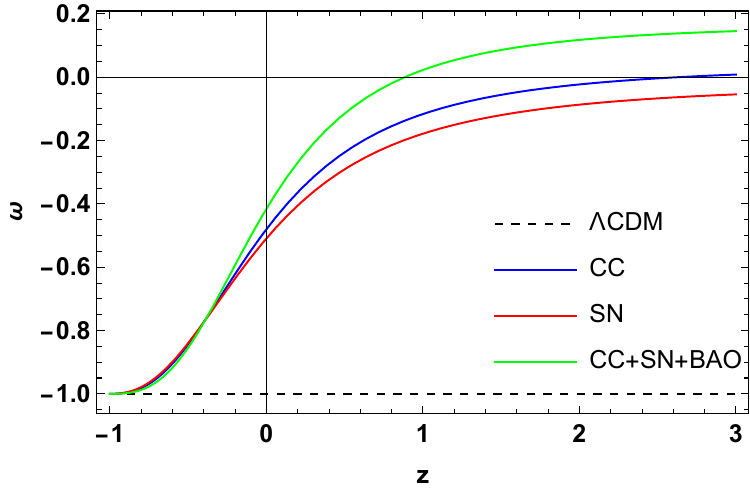}
\caption{Plot of EoS parameter $\omega$ versus redshift $z$.}
\label{F_w}
\end{figure}

We present plots of the matter–energy density and the EoS parameter against redshift $z$ in Figs. \ref{F_rho} and \ref{F_w}, corresponding to the constrained values of the model parameters. From Fig. \ref{F_rho}, we observe that the cosmic matter–energy density behaves as expected, showing a positive trend and diminishing as the universe expands towards the distant future. This behavior aligns with the standard expectations for the evolution of the universe. For all the considered values of the model parameters, the model begins in a matter-dominated era (at early times), transitions through the quintessence phase (at present), and ultimately approaches the $\Lambda$CDM model (at late times). Also, we observe that the present values of the EoS parameter exhibit quintessence-like behavior \cite{Koussour3,Koussour4,Koussour5}.

\section{$\omega-\omega'$ Analysis}\label{sec6}

The behavior of quintessence DE models has been analyzed using the $\omega - \omega'$ method, first proposed by Caldwell and Linder \cite{Caldwell_2005}, where prime represents a derivative with respect to $\ln a$, the natural logarithm of the scale factor. They investigated the limits of the quintessence model, distinguishing between thawing (where $\omega' > 0$ for $\omega < 0$) and freezing (where $\omega' < 0$ for $\omega < 0$) regions by constructing the $\omega - \omega'$ plane. Subsequently, this method has been utilized in other prominent dynamical DE models, including more general forms of quintessence \cite{Scherrer_2006}, phantom \cite{Chiba_2006}, quintom \cite{Guo_2006}, and various other models. The universe's expansion is more accelerated in the freezing region than in the thawing region. To analyze the $\omega - \omega'$ behavior for this model, we calculate $\omega'$ by differentiating Eq. (\ref{w}) with respect to $\ln a$, yielding:
\begin{equation}
    \omega'(z)=\frac{d\omega(z)}{d\ln a}=-(1+z)\frac{d\omega(z)}{dz}.
\end{equation}

Thus,
\begin{equation}
    \omega'(z)=-\frac{4 (2 \alpha -1) n^2 (1+z)^{2 n}}{3 \alpha  \left((1+z)^{2 n}+1\right)^2}.
\end{equation}

\begin{figure}[h]
\centering
\includegraphics[scale=0.7]{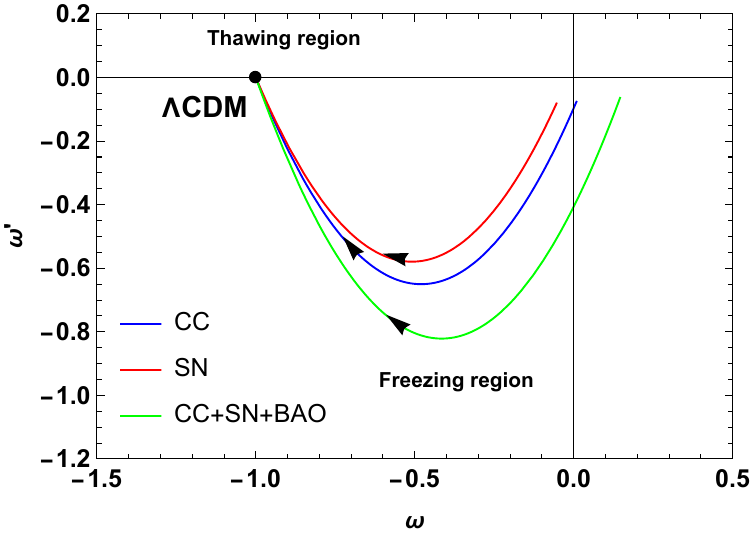}
\caption{Plot of $\omega - \omega'$ plane.}
\label{F_w-w'}
\end{figure}

We can construct the $\omega - \omega'$ plane by plotting $\omega'$ against $\omega$ for the constrained values of the model parameters, as depicted in Fig. \ref{F_w-w'}. It is evident that the $\Lambda$CDM limit ($\omega' = 0$ for $\omega = -1$) can be attained for the model. Also, we have identified only the freezing region in that plane because $\omega' < 0$ for $\omega < 0$. Therefore, there is no thawing region available in our model. This indicates that the plane analysis is in agreement with the accelerated expansion of the universe.

\section{Stability analysis}\label{sec7}

In linear perturbation theory, the square of the sound speed $\nu_{s}^2$ plays a crucial role. The stability or instability of a perturbed mode can be assessed by determining the sign of $\nu_{s}^2$. A positive sign (real value of $\nu_{s}^2$) indicates a periodic propagating mode for a density perturbation, indicating stability. Conversely, a negative sign (imaginary value of $\nu_{s}^2$) signifies an exponentially growing mode for a density perturbation, indicating instability \cite{Myung_2007,Kim_2008}. 
The square of the sound speed $\nu_{s}^2$ is defined as
\begin{equation}
    \nu_{s}^2=\frac{dp}{d\rho}=\frac{\dot p}{\dot \rho}=\frac{\rho}{\dot \rho} \dot \omega+\omega.
\end{equation}

Using Eqs. (\ref{F1}), (\ref{F2}), and (\ref{w}), the square of the sound speed can be obtained as
\begin{equation}
    \nu_{s}^2(z)=-1+\frac{2 (2 \alpha -1) n \left(\alpha +(1+z)^{2 n}\right)}{3 \alpha  \left((1+z)^{2 n}+1\right)}.
\end{equation}

\begin{figure}[h]
\centering
\includegraphics[scale=0.7]{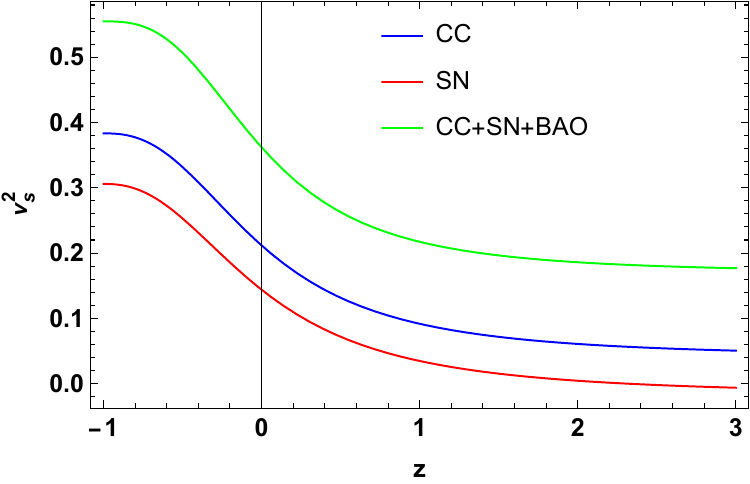}
\caption{Plot of sound speed $\nu_{s}^2$ versus redshift $z$.}
\label{F_vs}
\end{figure}

Fig. \ref{F_vs} shows that the square of the sound speed remains positive and $0<\nu_{s}^2<1$ throughout the cosmic evolution, indicating the stability of our model. This behavior is crucial for ensuring that density perturbations do not lead to instabilities in the system.

\section{Conclusion}\label{sec8}

Our analysis within the framework of $f(R, L_m)$ gravity theory \cite{Harko/2010} offers a comprehensive understanding of the freezing quintessence scenario in late-time cosmic expansion. We focused on a specific non-linear $f(R, L_m)$ model, $f(R,L_m)=\frac{R}{2}+L_m^\alpha$, where $\alpha$ is a free parameter \cite{Jaybhaye/2022,Myrzakulov/2023,Myrzakulova/2024,Koussour/2024}. A solution for this model is derived using an appropriate parametrization of the scale factor, given by $a(t) = \sqrt[n]{\sinh(t)}$, where $n > 0$ is an arbitrary constant \cite{Chawla/2012}. Moreover, we have successfully constrained the model parameters using a combination of the CC dataset, the recently published Pantheon+ (SN) dataset, as well as the combined CC+SN+BAO dataset, achieving results that are in excellent agreement with observational data. 

Furthermore, we have examined the behavior of the Hubble parameter, deceleration parameter, matter-energy density, and the EoS parameter for the constrained values of the model parameters. Fig. \ref{H_err} indicates that the Hubble parameter decreases with the evolution of the universe, aligning with observational findings. Specifically, we observed that $H$ increases as $z$ increases. In addition, we found that the present values of the Hubble parameter are consistent with those reported by the Planck collaboration ($H_0 = 67.4 \pm 0.5$ km s$^{-1}$ Mpc$^{-1}$). During the early stages of evolution, the universe was predominantly matter-dominated, leading to a decelerating phase. However, as the universe expanded, the phase transitioned from deceleration to acceleration, primarily driven by the dominance of DE. We have investigated the deceleration parameter and depicted its variation with respect to the redshift in Fig. \ref{F_q}. It is evident from this plot that the deceleration parameter's sign changed at a transition redshift of approximately 0.3-1, indicating the universe's transition from a decelerating phase to an accelerating phase. This transition serves as a strong indication of the physical validity of our model. The EoS parameter and matter-energy density follow expected trends over cosmic time, transitioning through different phases and approaching the $\Lambda$CDM model at late times (see Figs. \ref{F_rho} and \ref{F_w}). The present values
of the EoS parameter exhibit quintessence-like behavior. Importantly, the model demonstrates consistency with the $\Lambda$CDM limit and exhibits freezing behavior in the $\omega - \omega'$ plane (see Fig. \ref{F_w-w'}). In the freezing region, the universe experiences a more accelerated expansion compared to the thawing region. Also, the sound speed remains within stable limits throughout cosmic evolution, ensuring the stability of the model against density perturbations (see Fig. \ref{F_vs}).

In conclusion, we have derived a set of physically viable solutions within the framework of the non-linear $f(R, L_m)$ model. Our analysis of various cosmological parameters indicates the stability of our model, affirming that the modified $f(R, L_m)$ gravity is a credible approach for describing the accelerating phase of the current universe. While $f(R)$ gravity provides a robust framework for describing the entire cosmological history \cite{Com1,Com2}, our results demonstrate that the inclusion of a matter-geometry coupling in $f(R, L_m)$ gravity introduces additional degrees of freedom that significantly impact cosmic evolution. These differences manifest in the non-conservation of the energy-momentum tensor and the emergence of an additional force, leading to deviations from geodesic motion. Furthermore, the $f(R, L_m)$ framework offers novel thermodynamic perspectives, such as irreversible matter creation and enhanced entropy production. This comparative analysis underscores the potential of $f(R, L_m)$ gravity to provide a more comprehensive understanding of the interplay between matter and geometry, while also highlighting the need for further observational and theoretical studies to constrain its parameter space and validate its cosmological implications.

\section*{Acknowledgment}
This research was funded by the Science Committee of the Ministry of Science and Higher Education of the Republic of Kazakhstan (Grant No. AP22682760).

\section*{Data Availability Statement}
There are no new data associated with this article.


\end{document}